\documentclass[aps,superscriptaddress,reprint]{revtex4-1}
\usepackage{amsmath}
\usepackage{amssymb}
\usepackage{graphicx}
\usepackage[colorlinks,citecolor=blue, linkcolor=blue,hyperindex,CJKbookmarks,dvipdfm]{hyperref}

\begin{document}

\title{Generation of single entangled photon-phonon pairs\\
 via an atom-photon-phonon interaction}
\author{Xun-Wei Xu}
\email{davidxu0816@163.com}
\affiliation{Department of Applied Physics, East China Jiaotong University, Nanchang,
330013, China}
\author{Hai-Quan Shi}
\affiliation{School of Materials Science and Engineering, Nanchang University, Nanchang
330031, China}
\affiliation{Department of Applied Physics, East China Jiaotong University, Nanchang,
330013, China}
\author{Jie-Qiao Liao}
\affiliation{Key Laboratory of Low-Dimensional Quantum Structures and Quantum Control of
Ministry of Education, Department of Physics and Synergetic Innovation
Center for Quantum Effects and Applications, Hunan Normal University,
Changsha 410081, China}
\author{Ai-Xi Chen}
\email{aixichen@zstu.edu.cn}
\affiliation{Department of Physics, Zhejiang Sci-Tech University, Hangzhou 310018, China}
\affiliation{Department of Applied Physics, East China Jiaotong University, Nanchang,
330013, China}
\date{\today }

\begin{abstract}
Quantum blockade and entanglement play important roles in quantum
information and quantum communication as quantum blockade is an effective
mechanism to generate single photons (phonons) and entanglement is a crucial
resource for quantum information processing. In this work, we propose a method to
generate single entangled photon-phonon pairs in a hybrid optomechanical
system. We show that photon blockade, phonon blockade, and photon-phonon correlation
and entanglement can be observed via the atom-photon-phonon
(tripartite) interaction, under the resonant atomic driving. The correlated and
entangled single photons and single phonons, i.e., single entangled photon-phonon pairs, can be generated in both the
weak and strong tripartite interaction regimes. Our results may have
important applications in the development of highly complex quantum networks.
\end{abstract}

\maketitle

%\email{44175330@qq.com}

%\email{jqliao@hunnu.edu.cn}

\section{Introduction}

Optomechanical systems with parametric coupling between optical and mechanical
modes provide us a perfect platform for manipulating the states of photons and phonons~\cite{AspelmeyerARX13}. As an important application, photon (phonon)
blockade~\cite{ImamogluPRL97,BirnbaumNat05,YXLiuPRA10}, that only allows single photon (phonon) excitation in the optical
(mechanical) mode, based on optomechanical interaction has attracted
significant interest in the past few years. A number of designs based on
diverse mechanisms are proposed to demonstrate photon (phonon) blockade in
optomechanical systems, such as photon (phonon) blockade based on strong
optomechanical couplings~\cite%
{RablPRL11,NunnenkampPRL11,StannigelPRL12,XWXuPRA13a,KronwaldPRA13,JQLiaoPRA13,XYLuPRL15,DHuPRA15,HXiePRA16,SeokPRA17,HXiePRA17}
and photon (phonon) blockade in weak nonlinear regime induced by quantum
interference~\cite{XWXuJPB13,SavonaArx13,HQShiSR18,MWangPRA19,BJLiPR19}.

In a recent experiment~\cite{RiedingerNat16}, the non-classical correlations between single photons and phonons from a nanomechanical resonator was reported by driving the nanomechanical photonic crystal cavity with blue-detuned optical pulse.
After that, we studied the photon and phonon
statistics in a quadratically coupled optomechanical system, and show that
both photon blockade and phonon blockade can be observed in the same
parameter regime, and more important, the single photons and single phonons
are strongly anticorrelated~\cite{XWXuPRA18}. Here, we will do a further study
and propose a method to generate correlated single photons and single
phonons under the constant atomic driving. Even more interestingly, we will show that the correlated single photons
and single phonons are entangled with each other, i.e., they are single entangled
photon-phonon pairs.

Entangled states have great significance of both fundamental physics study
and applications in quantum information processing and quantum
communication. The optomechanical entanglement has already been
proposed theoretically~\cite%
{VitaliPRL07,HartmannPRL08,BorkjePRL11,BarzanjehPRL12,YDWangPRL12,LTianPRL12}
and demonstrated experimentally~\cite%
{PalomakiSci13,RiedingerNat18,Ockeloen-KorppiNat18,MarinkovicPRL18}.
Optomechanical systems provide a perfect platform to generate both bipartite~%
\cite{WJNiePRA12,XWXuPRA13,JQLiaoPRA14,XYLuPRA18} and multipartite~\cite%
{PaternostroPRL07,GenesPRA08,HTTanPRA11,XuerebPRA12,JLiPRL18} entanglement.
However, there are substantial differences between the entanglement we will
discuss in this paper and entanglement proposed before. One striking
difference is the entanglement we proposed here is for single photons and
phonons, which is non-Gaussian, so that the generally adopted method, i.e., the linearization of
the optomechanical interaction, is no longer applicable.

Inspired by a recent experiment~\cite{FTianArx19}, in which the coupling between an optomechanical resonator with
two-level emitters was realized, here we consider a hybrid system which
enables a tripartite interaction between a two-level atom, an optical mode,
and a mechanical mode~\cite{YChangJPB09}. We study the generation of single entangled
photon-phonon pairs, which are uesful for quantum information and quantum communication. Such
atom-photon-phonon interactions were proposed to provide an optically
controllable interaction between a two-level atom and a macroscopic
mechanical oscillator~\cite{CotrufoPRL17,MWangPRA19} by driving the optical
mode strongly. Nevertheless, in this paper we drive the two-level atom
coherently and show
that single entangled photon-phonon pairs can be generated in the hybrid optomechanical system. The single entangled photon-phonon pairs have potential application in the development of highly complex quantum networks.

The remainder of this paper is organized as follows. In Sec.~II, we
introduce the theoretical model of a hybrid optomechanical system, and show
the simple derivation of the atom-photon-phonon interaction and the energy spectrum of the Hamiltonian. In
Sec.~III, the photon and phonon statistics, and the quantum correlation
between the photons and phonons are discussed numerically. Finally, a
summary is given in Sec.~IV.

\section{Model and Hamiltonian}

\begin{widetext}
\begin{figure*}[tbp]
\includegraphics[bb=56 292 543 609, width=12 cm, clip]{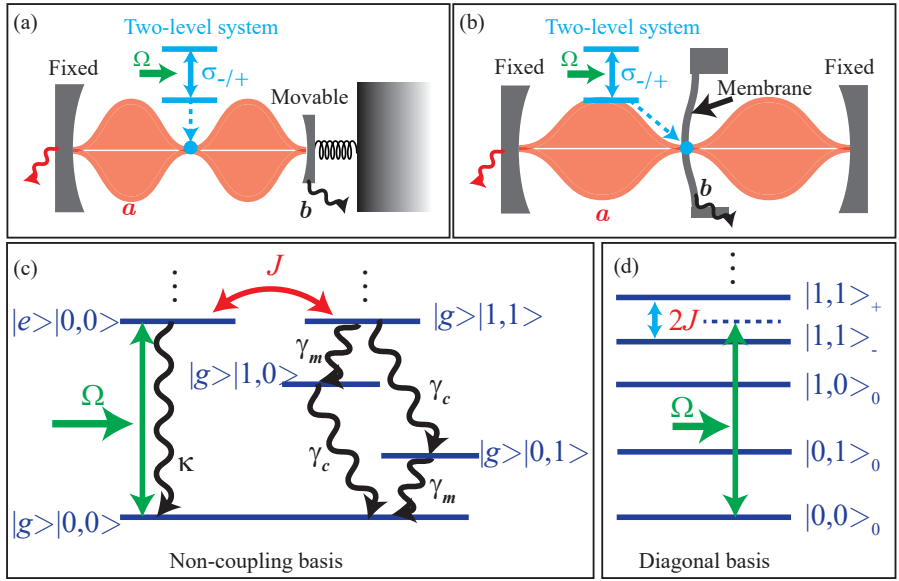}
\caption{(Color online) Schematics of the hybrid systems with atom-photon-phonon
interactions: (a) a two-level atom in an optical cavity with a movable end
mirror: (b) a two-level atom imbedded in a membrane inside an optical cavity.
The energy spectrum of the hybrid optomechanical system [see the Hamiltonian in Eq.~(%
\protect\ref{Eq5})] given (c) in the non-coupling basis and (d) in the
diagonal basis.
}
\label{fig1}
\end{figure*}
\end{widetext}

We study a hybrid system with a two-level atom ($\sigma _{\pm}$ being the
ladder operators) of transition frequency $\omega _{0}$ and a mechanical
resonator $b$ of resonance frequency $\omega _{m}$ in an optical cavity $a$ of
resonance frequency $\omega _{c}$, as shown in Fig.~\ref{fig1}(a) and (b). Here, we
consider a special situation in which the mechanical displacement $x$
induces a variation of the spatial distribution of the cavity field~\cite%
{CotrufoPRL17}, while the mechanical effect on the optical frequency $\omega _{c}$ can
be neglected. Thus, the coupling strength $g\left( x\right) $ between the
two-level atom and the optical mode depends on the position of the
mechiancal resonator $x$, which is described by the interaction Hamiltonian
under the rotating-wave approximation as ($\hbar =1$)%
\begin{equation}
H_{\mathrm{int}}=g\left( x\right) \left( \sigma _{+}a+\sigma _{-}a^{\dag
}\right) .
\end{equation}%

Typically the mechanical displacement $x$ is very small, and $g\left(
x\right) $ can be expanded to the first order in $x$,%
\begin{equation}
g\left( x\right) =g\left( 0\right) +J\left( b^{\dag }+b\right),
\end{equation}%
where $J\equiv \left( \partial g/\partial x\right) |_{x=0}x_{\mathrm{zpf}}$
is the tripartite atom-photon-phonon interaction strength. In the specific
condition that the two-level atom is placed at the node of the optical mode
with mechanical resonator in equilibrium, i.e., $g\left( 0\right) =0$, and
the only possible interaction between them is the atom-photon-phonon
interaction as%
\begin{equation}
H_{\mathrm{int}}=J\left( b^{\dag }+b\right) \left( \sigma _{+}a+\sigma
_{-}a^{\dag }\right) .
\end{equation}%
Under particular resonant conditions, the tripartite interaction allows
swapping the excitation between the three quantum systems. Under the
conditions $\omega _{0}=\omega_{c}+\omega _{m}$ and $\min\{\omega _{0},\omega_{c}\}\gg \omega _{m}\gg J$, the
Hamiltonian of the resonant interaction reads
\begin{equation}
H_{\mathrm{int}}=J\left( \sigma _{+}ab+\sigma _{-}a^{\dag }b^{\dag }\right) ,
\end{equation}%
which describes the simultaneous generation of a photon and a phonon with
the two-level atom jumping from the excited state to its ground state and the reverse process. This
tripartite interaction provides us an effective way to generate photon-phonon
pairs. Such a hybrid system can be implemented in the electromechanical
systems~\cite{TeufelNat11,MasselNC12,PalomakiSci13,SuhSci14} with artificial
atom at the node or in a Fabry-P\'{e}rot cavity with a membrane containing
two-level atoms in the node of the cavity mode~\cite%
{ThompsonNat08,Flowers-JacobsAPL12,HXuNat16,HXuNC17}.

Next, we consider the case that the two-level atom is pumped by a coherent field (strength
$\Omega $, frequency $\omega _{p}$), and the total Hamiltonian\ for the
hybrid system in the rotating frame with respect to $R\left( t\right) =\exp %
\left[ i\omega _{p}\sigma _{+}\sigma _{-}t+i\left( \omega _{p}-\omega
_{m}\right) a^{\dag }at+i\omega _{m}b^{\dag }bt\right] $ reads
\begin{equation}  \label{Eq5}
H=\Delta \sigma _{+}\sigma _{-}+\Delta a^{\dag }a +J\left( \sigma _{+}ab+\sigma _{-}a^{\dag }b^{\dag }\right)+\Omega \sigma_{x},
\end{equation}
where we introduce the detuning $\Delta \equiv \omega _{0}-\omega _{p}=\omega _{c}+\omega
_{m}-\omega _{p}$.

The energy spectrum of the Hamiltonian in Eq.~(\ref{Eq5}) for hybrid
optomechanical system is shown in Figs.~\ref{fig1}(c) and \ref{fig1}(d). In the
non-coupling basis [Fig.~\ref{fig1}(c)], $\left\vert e\right\rangle $ ($%
\left\vert g \right\rangle $) denotes the excited (ground) state of the
two-level atom, and $\left\vert n,m\right\rangle $ represents the Fock
state with $n$ photons in the optical mode and $m$ phonons in the mechanical
mode. In Fig.~\ref{fig1}(d), we denote the eigenstates in the diagonal basis as $\left\vert 0,0\right\rangle_{0}
\equiv \left\vert g\right\rangle\left\vert 0,0\right\rangle $, $\left\vert
1,0\right\rangle_{0} \equiv \left\vert g\right\rangle \left\vert
1,0\right\rangle $, $\left\vert 0,1\right\rangle_{0} \equiv \left\vert
g\right\rangle \left\vert 0,1\right\rangle $, $\left\vert
1,1\right\rangle_{\pm } \equiv (\left\vert g\right\rangle\left\vert
1,1\right\rangle \pm \left\vert e\right\rangle\left\vert 0,0\right\rangle)/%
\sqrt{2} $ with eigenvalues $0$, $\omega_c$, $\omega_m$, $\omega_0\pm J$, respectively.

\section{Correlation and entanglement}

\begin{widetext}
\begin{figure*}[tbp]
\includegraphics[bb=29 433 541 632, width=16 cm, clip]{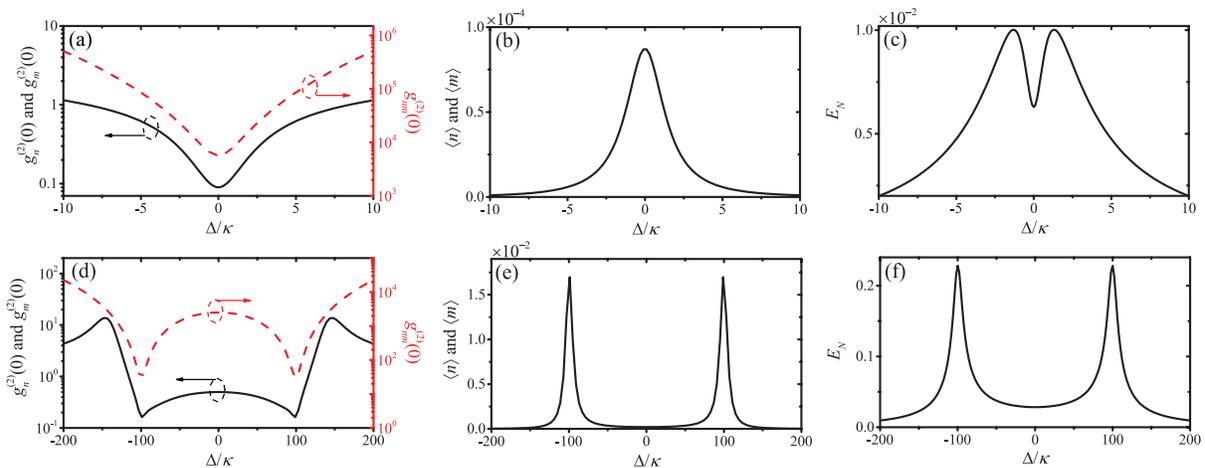}
\caption{(Color online) In panels (a) and (d), the equal-time second-order correlation
functions [$g^{(2)}_{n}(0)$ and $g^{(2)}_{m}(0)$] and cross-correlation
function $g^{(2)}_{nm}(0)$ are plotted as functions of the detuning $\Delta/%
\protect\kappa$. In panels (b) and (e), the mean photon (phonon) number [$\langle n
\rangle=\langle m \rangle$] is plotted as a function of
the detuning $\Delta/\protect\kappa$. In panels (c) and (f), the logarithmic negativity $E_N$ is plotted as a function of the detuning $\Delta/\protect\kappa$. We set $J=0.1 \protect\kappa$ in (a)
and (b) and set $J=100 \protect\kappa$ in (c) and (d). Other used parameters
are $\protect\gamma_c=\protect\gamma_m=10 \protect\kappa$, $\Omega=\protect%
\kappa$, and $m_{\mathrm{th}}=0$.}
\label{fig2}
\end{figure*}
\end{widetext}

To quantify the statistics of the phonons and photons in the system, we
consider the equal-time second-order correlation functions $g^{(2)}_{n}(0)$ and $g^{(2)}_{m}(0)$, and cross-correlation
function $g^{(2)}_{nm}(0)$ in the steady
state ($t\rightarrow \infty $) defined by
\begin{equation}
g_{n}^{\left( 2\right) }\left( 0\right) \equiv \frac{\left\langle a^{\dag
}a^{\dag }aa\right\rangle }{\langle n\rangle^{2}},
\end{equation}%
\begin{equation}
g_{m}^{\left( 2\right) }\left( 0\right) \equiv \frac{\left\langle b^{\dag
}b^{\dag }bb\right\rangle }{\langle m\rangle^{2}},
\end{equation}%
\begin{equation}
g_{nm}^{\left( 2\right) }\left( 0\right) \equiv \frac{\left\langle a^{\dag
}b^{\dag }ba\right\rangle }{\langle n\rangle \langle m\rangle},
\end{equation}%
where $\langle n\rangle\equiv \left\langle a^{\dag }a\right\rangle $ and $\langle m\rangle\equiv
\left\langle b^{\dag }b\right\rangle $ are the mean photon and phonon
numbers. The dynamic behavior of the total open system is described by the
master equation for the density matrix $\rho $ of the system~\cite%
{Carmichael93}%
\begin{eqnarray}  \label{Eq9}
\frac{\partial \rho }{\partial t} &=&-i\left[ H,\rho \right] +\kappa
L[\sigma _{-}]\rho +\gamma _{c}L[a]\rho  \notag  \label{Eq6} \\
&&+\gamma _{m}\left( m_{\mathrm{th}}+1\right) L[b]\rho +\gamma _{m}m_{%
\mathrm{th}}L[b^{\dag }]\rho ,
\end{eqnarray}%
where $L[o]\rho =o\rho o^{\dag }-\left( o^{\dag }o\rho +\rho o^{\dag
}o\right) /2$ denotes a Lindbland term for an operator $o$; $\kappa $ is the
damping rate of the two-level atom and $\gamma _{c}$ ($\gamma _{m}$) is the
damping rate of the optical (mechanical) mode; $m_{\mathrm{th}}$ is the mean
thermal phonon number. We assume that the frequencies of the two-level atom and
the optical mode are so high that the thermal effect can be neglected.

The equal-time second-order correlation functions [$g^{(2)}_{n}(0)$ and $%
g^{(2)}_{m}(0)$] and cross-correlation function $g^{(2)}_{nm}(0)$ are
plotted as functions of the detuning $\Delta/\kappa$ in Fig.~\ref{fig2}
under both weak-coupling condition [(a) $J=0.1 \kappa$] and strong-coupling
condition [(c) $J=100 \kappa$]. It is clear that photon blockade and phonon
blockade, i.e., $g^{(2)}_{n}(0)=g^{(2)}_{m}(0)<1$, appear simultaneously around $|
\Delta| =J$ for the same parameters. Simultaneously, the single photons and single
phonons generated by photon blockade and phonon blockade are strongly
correlated with each other, i.e., $g^{(2)}_{nm}(0)\gg 1$. The optimal
detuning $\Delta$ for correlated photon blockade and phonon blockade depends on the coupling strength $J$: $\Delta=0$ for weak coupling and $|
\Delta| \approx J$ for strong coupling. Moreover, the mean photon (phonon)
number $\langle n \rangle=\langle m \rangle$ in the weak-coupling case
is much smaller than the one in the strong-coupling case.

Physically, the single photon and phonon pairs are generated one by one with the two-level atom jumping from the excited state to its ground state. In the weak-coupling regime ($J\ll \kappa$), the system is driven resonantly with detuning $\Delta=0$ because the states $\left\vert
1,1\right\rangle_{+}$ and $\left\vert
1,1\right\rangle_{-}$ are not resolved. In the strong-coupling regime ($J\gg \kappa$), the system should be investigated by the dressed states as shown in Fig.~\ref{fig1}(d), and the single photon and phonon pairs are generated with detuning $\Delta = \pm J$ for resonant pumping.

In order to understand the behavior of the cross-correlation function $g^{(2)}_{nm}(0)$, we can give the expression of $g^{(2)}_{nm}(0)$ approximately.
Under the weak-exciting condition, i.e., $\max\{\langle n\rangle, \langle m\rangle\}\ll 1$, we have mean photon (phonon) number
\begin{equation}
\langle n\rangle \approx \rho_{5,5}+\rho_{4,4},
\end{equation}
\begin{equation}
\langle m\rangle \approx \rho_{5,5}+\rho_{3,3},
\end{equation}
and the cross-correlation function $g^{(2)}_{nm}(0)$
\begin{equation}
g^{(2)}_{nm}(0) \approx \frac{\rho_{5,5}}{(\rho_{5,5}+\rho_{3,3})(\rho_{5,5}+\rho_{4,4})},
\end{equation}
where $\rho_{3,3}=\langle g|\langle 0,1|\rho |g\rangle|0,1\rangle$, $\rho_{4,4}=\langle g|\langle 1,0|\rho |g\rangle|1,0\rangle$, and $\rho_{5,5}=\langle g|\langle 1,1|\rho |g\rangle|1,1\rangle$, and they satisfy the relations
\begin{equation}\label{Eq13}
\rho_{3,3} \approx \frac{\gamma_c}{\gamma_m} \rho_{5,5},
\end{equation}
\begin{equation}\label{Eq14}
\rho_{4,4} \approx \frac{\gamma_m}{\gamma_c} \rho_{5,5}.
\end{equation}
If we set $\gamma_c=\gamma_m$, then we have $\rho_{3,3} \approx\rho_{4,4}  \approx\rho_{5,5}$, $\langle n\rangle=\langle m\rangle \approx 2\rho_{5,5}$, and
\begin{equation}\label{Eq15}
g^{(2)}_{nm}(0)\approx  \frac{1}{2\langle n\rangle}.
\end{equation}
Under the resonant conditions at the detuning $|\Delta|=J$, we have maximum $\langle n\rangle=\langle m\rangle$, and thus minimum cross-correlation function $g^{(2)}_{nm}(0)$, corresponding to the dips around the detuning $|\Delta|=J$.

\begin{figure}[tbp]
\includegraphics[bb=55 225 540 619, width=8.5 cm, clip]{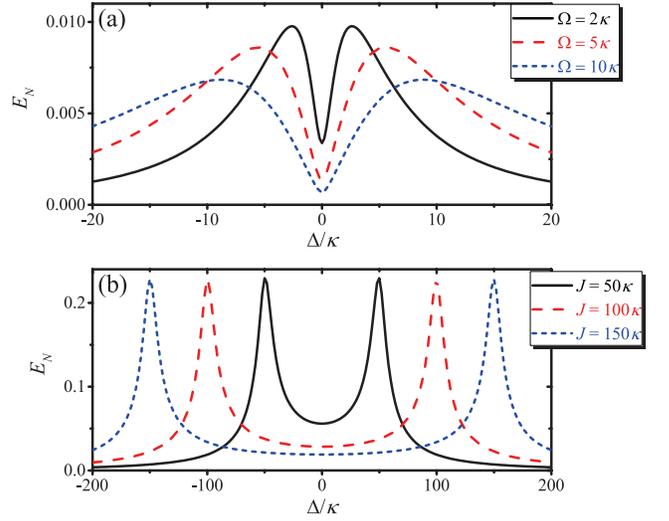}
\caption{(Color online) The logarithmic negativity $E_N$ is plotted as a function of the detuning $\Delta/\protect\kappa$ for different driving strength $\Omega$ in (a) and for different coupling strength $J$ in (b). We set $J=0.1 \protect\kappa$ in (a) and set $\Omega=\kappa$ in (b). Other used parameters
are $\gamma_c=\protect\gamma_m=10\kappa$ and $m_{\mathrm{th}}=0$.}
\label{fig3}
\end{figure}

\begin{figure}[tbp]
\includegraphics[bb=68 217 517 619, width=7.5 cm, clip]{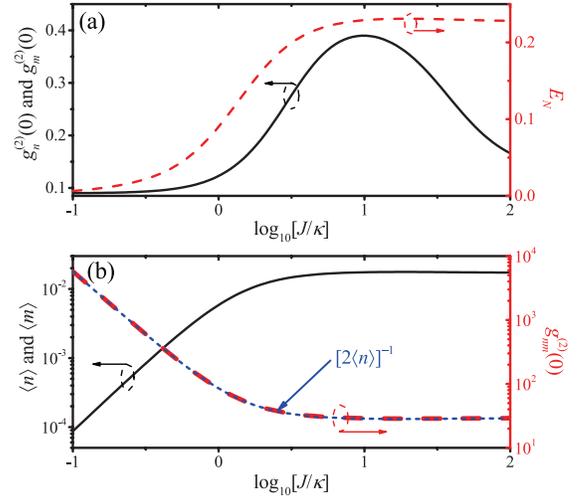}
\caption{(Color online) (a) The equal-time second-order correlation
functions [$g^{(2)}_{n}(0)$ and $g^{(2)}_{m}(0)$] the logarithmic negativity $%
E_{N}$ are plotted as functions of the coupling strength
$\log_{10} [J/\protect\kappa]$. (b) The mean photon (phonon) number [$%
\langle n \rangle$ and $\langle m \rangle$] and cross-correlation
function $g^{(2)}_{nm}(0)$ are plotted as functions of $\log_{10} [J/%
\protect\kappa]$. There are two curves for $g^{(2)}_{nm}(0)$, where the red dashed one is obtained from Eq.~(\ref{Eq9}) and the blue short-dashed one is obtained from Eq.~(\ref{Eq15}). Other used parameters are $|\Delta|=J$, $\protect\gamma_c=%
\protect\gamma_m=10 \protect\kappa$, $\Omega=\protect\kappa$, and $m_{%
\mathrm{th}}=0$.}
\label{fig4}
\end{figure}

It's not hard to guess that the strongly correlated single photons and
single phonons generated by photon blockade and phonon blockade are
entangled with each other. The entanglement between the optical and
mechanical modes can be characterized by the logarithmic negativity~\cite%
{VidalPRA02}
\begin{equation}
E_{N}=\log _{2}\left\Vert \rho _{AB}^{T_{A}}\right\Vert_1 ,  \label{Eq10}
\end{equation}%
where the symbol $\left\Vert \cdot \right\Vert_1 $ denotes the trace norm,
and $\rho _{AB}^{T_{A}}$ is the partial transpose of the reduced density
matrix $\rho _{AB}$ of the optical and mechanical modes.
It is worth mentioning that the entangled state for the single photons and single
phonons obtained here is non-Gaussian. Thus the logarithmic negativity for Gaussian states~\cite{AdessoPRA04} widely used in the previous works~
\cite{VitaliPRL07,HartmannPRL08,BorkjePRL11,BarzanjehPRL12,YDWangPRL12,LTianPRL12} cannot be used to accurately describe the entangled state here.

The logarithmic negativity $E_{N}$ is shown in Figs.~\ref{fig2}(c) and \ref{fig2}(f).
Obviously, the strongly correlated single photons and single phonons
generated by photon blockade and phonon blockade are entangled with each
other in both the weak ($J<\kappa $) and strong ($J>\kappa $) coupling
regimes. In the weak-coupling regime as shown in Fig.~\ref{fig2}(c), there is a dip around the detuning $%
\Delta =0$, which is induced by the quantum interferences between two
routes: (a) the direct transition channel $\left\vert g\right\rangle \left\vert 0,0\right\rangle \overset{%
\Omega }{\rightarrow }\left\vert e\right\rangle \left\vert
0,0\right\rangle \overset{J}{\rightarrow }\left\vert g\right\rangle
\left\vert 1,1\right\rangle $; (b) the indirect transition channel $%
\left\vert g\right\rangle \left\vert 0,0\right\rangle \overset{%
\Omega }{\rightarrow }\left\vert e\right\rangle \left\vert 0,0\right\rangle \overset{\Omega }{%
\rightarrow }\left\vert g\right\rangle \left\vert 0,0\right\rangle \overset{%
\Omega }{\rightarrow }\left\vert e\right\rangle \left\vert 0,0\right\rangle
\overset{J}{\rightarrow }\left\vert g\right\rangle \left\vert
1,1\right\rangle $ (or higher-order variants). Thus the width of the dip depends on the driving strength $\Omega$, as shown in Fig.~\ref{fig3}(a). Similar mechanism can induce transparency in lambda-type three-level atoms~\cite{HarrisPT97,FleischhauerRMP05} and optomechanical systems~\cite%
{AgarwalPRA10,WeisSci10,SafaviNaeiniNat11}.
Differently, in Fig.~\ref{fig2}(f), there are two peaks around the detunings $\Delta = \pm J$ in the strong-coupling regime. This phenomenon can be understood by analyzing the energy spectrum shown in Fig.~\ref{fig1}(d): the transition process $\left\vert 0,0\right\rangle_{0}\rightarrow \left\vert 1,1\right\rangle_{\pm } $ is resonantly enhanced with detunings $\Delta = \pm J$. As a consequence, we can shift the optimal value of the detuning for entanglement by tuning
the coupling strength $J$ as shown in Fig.~\ref{fig3}(b).

\begin{figure}[tbp]
\includegraphics[bb=52 196 563 634, width=8.5 cm, clip]{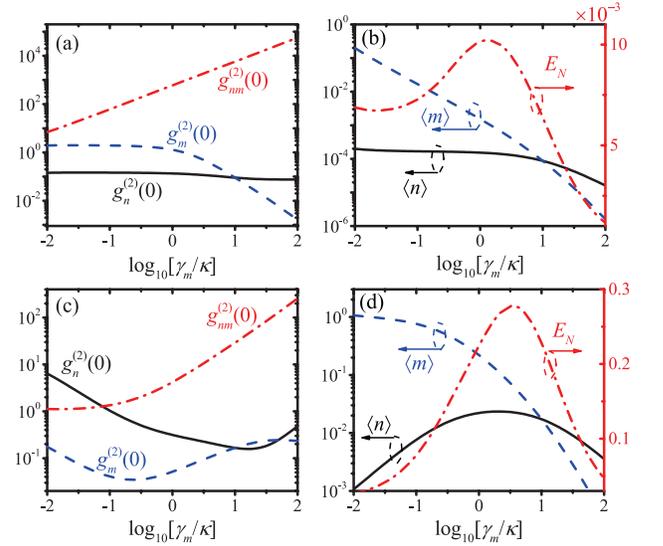}
\caption{(Color online) (a) and (c) the equal-time second-order correlation
functions [$g^{(2)}_{n}(0)$ and $g^{(2)}_{m}(0)$] and cross-correlation
function $g^{(2)}_{nm}(0)$ are plotted as functions of the mechanical
damping rate $\log_{10} [\protect\gamma_m/\protect\kappa]$. (b) and (d), the
mean photon (phonon) number [$\langle n \rangle$ and $\langle m \rangle$]
and the logarithmic negativity $E_{N}$ are plotted as functions of the
mechanical damping rate $\log_{10} [\protect\gamma_m/\protect\kappa]$. We
set $J=0.1 \protect\kappa$ in (a) and (b) and set $J=100 \protect\kappa$ in
(c) and (d). Other used parameters are $|\Delta|=J$, $\protect\gamma_c=10
\protect\kappa$, $\Omega=\protect\kappa$, and $m_{\mathrm{th}}=0$.}
\label{fig5}
\end{figure}

\begin{figure}[tbp]
\includegraphics[bb=52 272 508 671, width=8.5 cm, clip]{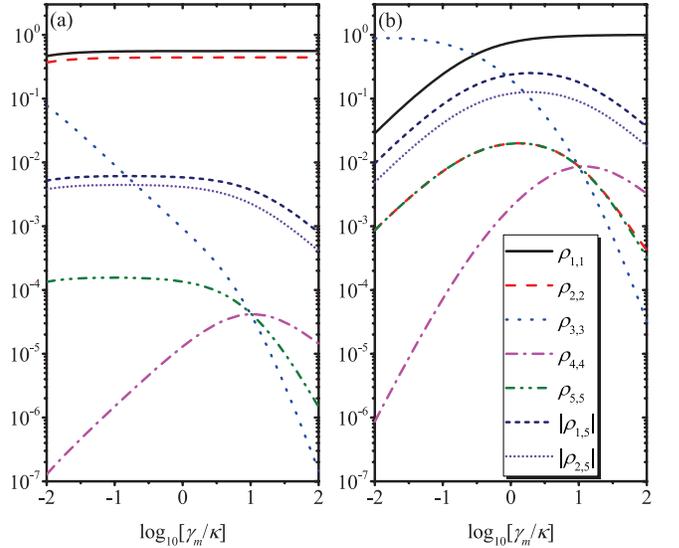}
\caption{(Color online) The elements of the density matrix $\rho $ from Eq.~(\ref{Eq9}) in the steady state are plotted as functions of the
mechanical damping rate $\log_{10} [\protect\gamma_m/\protect\kappa]$, where $\rho_{1,1}=\langle g|\langle 0,0|\rho |g\rangle|0,0\rangle$, $\rho_{2,2}=\langle e|\langle 0,0|\rho |e\rangle|0,0\rangle$, $\rho_{3,3}=\langle g|\langle 0,1|\rho |g\rangle|0,1\rangle$, $\rho_{4,4}=\langle g|\langle 1,0|\rho |g\rangle|1,0\rangle$, $\rho_{5,5}=\langle g|\langle 1,1|\rho |g\rangle|1,1\rangle$, $\rho_{1,5}=\langle g|\langle 0,0|\rho |g\rangle|1,1\rangle$, $\rho_{2,5}=\langle e|\langle 0,0|\rho |g\rangle|1,1\rangle$. We
set $J=0.1 \protect\kappa$ in (a) and set $J=100 \protect\kappa$ in
(b). Other used parameters are $|\Delta|=J$, $\protect\gamma_c=10
\protect\kappa$, $\Omega=\protect\kappa$, and $m_{\mathrm{th}}=0$.}
\label{fig6}
\end{figure}

Figure~\ref{fig4} shows the second-order correlation functions [$%
g^{(2)}_{n}(0)$ and $g^{(2)}_{m}(0)$] and cross-correlation function $%
g^{(2)}_{nm}(0)$ with the coupling strength $J$ from weak to strong.
The mean photon (phonon) number [$\langle n \rangle=\langle m \rangle$] and logarithmic negativity $E_{N}$
increase with the enhancing of the coupling strength $J$.
As shown in Fig.~\ref{fig4}(b), the cross-correlation function $g^{(2)}_{nm}(0)$ decreases with the increasing of the mean photon (phonon) number [$\langle n \rangle=\langle m \rangle$], and the numerical results (red dashed curve) agrees well with the analytical results given in Eq.~(\ref{Eq15}) (blue short-dashed curve).
The second-order correlation functions [$%
g^{(2)}_{n}(0)$ and $g^{(2)}_{m}(0)$] increases first with the mean photon (phonon) number, and then decreases with the coupling strength $J$, as the excitations of states $\left\vert
2,1\right\rangle_{\pm } \equiv (\left\vert g\right\rangle\left\vert
2,1\right\rangle \pm \left\vert e\right\rangle\left\vert 1,0\right\rangle)/%
\sqrt{2} $ and $\left\vert
1,2\right\rangle_{\pm } \equiv (\left\vert g\right\rangle\left\vert
1,2\right\rangle \pm \left\vert e\right\rangle\left\vert 0,1\right\rangle)/%
\sqrt{2} $ are suppressed for the enhancement of the effective damping rates with the coupling strength $J$.
The suitable coupling strength $J$ for observing correlated single photons and
single phonons, i.e., $g^{(2)}_{n}(0)=g^{(2)}_{m}(0)\ll 1$ and $%
g^{(2)}_{nm}(0)\gg 1$, is $J \ll \kappa$ or $J \gg \kappa$.

\begin{figure}[tbp]
\includegraphics[bb=52 196 563 634, width=8.5 cm, clip]{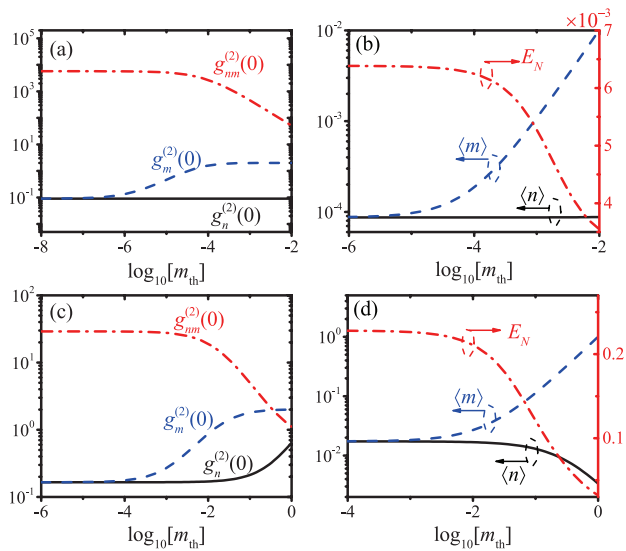}
\caption{(Color online) (a) and (c) the equal-time second-order correlation
functions [$g^{(2)}_{n}(0)$ and $g^{(2)}_{m}(0)$] and cross-correlation
function $g^{(2)}_{nm}(0)$ are plotted as functions of the mean thermal
phonon number $\log_{10} [m_{\mathrm{th}}]$. (b) and (d), the mean photon
(phonon) number [$\langle n \rangle$ and $\langle m \rangle$] and the
logarithmic negativity $E_{N}$ are plotted as functions of the mean thermal
phonon number $\log_{10} [m_{\mathrm{th}}]$. We set $J=0.1 \protect\kappa$
in (a) and (b) and set $J=100 \protect\kappa$ in (c) and (d). Other used
parameters are $|\Delta|=J$, $\protect\gamma_c=\protect\gamma_m=10 \protect%
\kappa$, and $\Omega=\protect\kappa$.}
\label{fig7}
\end{figure}

Generally, the damping rate of the mechanical mode is much smaller than the
damping rate of the optical mode. However, the effective damping of the
mechanical mode can be controlled and significantly enhanced by coupling the
mechanical mode to an auxiliary optical mode~\cite%
{WilsonRaePRL07,MarquardtPRL07,LiYPRB08,XWXuPRA15}. In addition, the phonon
statistics can be observed indirectly by measuring statistics of the photons
output from the auxiliary optical mode~\cite%
{DidierPRB11,RamosPRL13,CohenNat15,XWXuPRA16}. The dependence of the
second-order correlation functions [$g^{(2)}_{n}(0)$ and $g^{(2)}_{m}(0)$]
and cross-correlation function $g^{(2)}_{nm}(0)$ on the mechanical damping
rate $\gamma_m$ is shown in Fig.~\ref{fig5}. In the weak-coupling case
[Figs.~\ref{fig5}(a) and \ref{fig5}(b)], the correlation and
cross-correlation functions change monotonically with the mechanical damping
rate. In the strong-coupling case [Figs.~\ref{fig5}(c)
and \ref{fig5}(d)], the correlation and cross-correlation functions changing
non-monotonously with the mechanical damping rate.
The mean phonon number
$\langle m \rangle$ decreases rapidly with the mechanical damping rate in both weak- and strong-coupling regimes and
the mean photon number $\langle n \rangle$ decreases monotonously for weak coupling ($J\ll \kappa$). While
$\langle n \rangle$ increases first and then decreases with the mechanical damping rate in the strong coupling regime ($J\gg\kappa$), i.e., we can enhance photon emission by increasing the mechanical damping rate when $\gamma_m<\kappa$.
Moreover, there is a optimal mechanical damping rate $\gamma_m$ for entanglement $E_N$ around the point $\gamma_m \approx 1.32 \kappa$ ($\gamma_m \approx 3.47 \kappa$) in the case of $J=0.1 \kappa$ ($J=100 \kappa$).

These interesting phenomena can be understood by the probability distribution in the bare states as shown in Fig.~\ref{fig6}, where $\rho_{1,1}=\langle g|\langle 0,0|\rho |g\rangle|0,0\rangle$, $\rho_{2,2}=\langle e|\langle 0,0|\rho |e\rangle|0,0\rangle$, $\rho_{3,3}=\langle g|\langle 0,1|\rho |g\rangle|0,1\rangle$, $\rho_{4,4}=\langle g|\langle 1,0|\rho |g\rangle|1,0\rangle$, $\rho_{5,5}=\langle g|\langle 1,1|\rho |g\rangle|1,1\rangle$, $\rho_{1,5}=\langle g|\langle 0,0|\rho |g\rangle|1,1\rangle$, $\rho_{2,5}=\langle e|\langle 0,0|\rho |g\rangle|1,1\rangle$. It is clear that we have $\rho_{3,3}\approx \rho_{4,4} \approx\rho_{5,5}$ around $\gamma_m=\gamma_c$, which is agree with Eqs.~(\ref{Eq13}) and (\ref{Eq14}). In the weak-coupling regime ($J\ll \kappa$) as shown in Fig.~\ref{fig6}(a), most of the probability is distributed in the states $|g\rangle|0,0\rangle$ and $|e\rangle|0,0\rangle$; the probability in states $|g\rangle|1,1\rangle$ (as well as the off-diagonal elements $|\rho_{1,4}|$ and $|\rho_{1,5}|$, which determine the entanglement $E_N$ between the photons and phonons) increases slowly in the regime of $\gamma_m < \kappa$, and decreases rapidly when $\gamma_m > \kappa$; the probability in single phonon state $|g\rangle|0,1\rangle$ (single photon state $|g\rangle|1,0\rangle$) decreases (increases) in the regime of $\gamma_m < \gamma_c$, and decreases rapidly when $\gamma_m > \gamma_c$. Differently, in the strong-coupling regime ($J \gg \kappa$) as shown in Fig.~\ref{fig6}(b), most of the probability ($87.5\%$) is distributed in the single phonon state $|g\rangle|0,1\rangle$ when $\gamma_m \ll \kappa$, as the damping rate of the state $|g\rangle|0,1\rangle$ is much smaller than the other states; the probability in the ground state $|g\rangle|0,0\rangle$ increase monotonously with the mechanical damping rate; the probability in states $|e\rangle|0,0\rangle$ and $|g\rangle|1,1\rangle$ are almost the same, i.e., $\rho_{2,2}\approx \rho_{5,5}$, and they (as well as the off-diagonal elements $|\rho_{1,4}|$ and $|\rho_{1,5}|$) increases first and then decreases with the mechanical damping rate, which is corresponding to the phenomena of photon emission and entanglement enhancing by increasing the mechanical damping rate when $\gamma_m < \kappa$.

The thermal effect of the mechanical mode on the statistic properties of the
generated photons and phonons are shown in Fig.~\ref{fig7}. It is clear that
the thermal phonons have a significant effect on the statistic properties of
the generated photons and phonons. As the mean phonon number is much larger
in the strong-coupling regime than the one in the weak-coupling regime, the
correlated and entangled photon blockade and phonon blockade in the strong-coupling regime
is more robust against the thermal noise than the one in the weak-coupling
regime.

\section{Conclusions}

In summary, we have studied the photon and phonon statistics, and the
quantum correlation between photons and phonons in a hybrid optomechanical system
including an atom-photon-phonon (tripartite) interaction. We have shown that both the photon and
phonon blockade can be observed in the same parameter area, and the
generated single photons and single phonons are correlated and entangled
with each other. Moreover, the single entangled photon-phonon pairs can be observed in both the weak and strong tripartite interaction regime.
The phonons with low-loss can be used for quantum memories, and photons are suitable for the
transmission of quantum information. The generated single entangled photon-phonon pairs will have applications in quantum communication and the hybrid optomechanical system can serve as quantum transducers in building hybrid quantum networks.
In addition, the basic mechanism of this work can be generalized to a nondegenerate two-photon Jaynes-Cummings model~\cite{SCGouPRA89,AshrafPRA92}, to generate entangled photon pairs with different frequency, such as entangled microwave-optical photon pairs~\cite{CZhongARX19}.

%\section{Acknowledgement}
\vskip 2pc \leftline{\bf Acknowledgement}

We thank Yan-Jun Zhao, Hui Wang, and Qiang Zheng for helpful discussions.
X.-W.X. was supported by the National Natural Science Foundation of China
(NSFC) under Grant No.~11604096. A.-X.C. is supported by NSFC under Grant No.
11775190. J.-Q.L. is supported in part by National Natural Science Foundation of China under Grants No.~11822501 and No.~11774087, and Natural Science Foundation of
Hunan Province, China under Grant No.~2017JJ1021.

\bibliographystyle{apsrev}
\bibliography{ref}

\end{document}